\documentclass[11pt,a4paper]{article}
\pdfoutput=1
\usepackage[utf8]{inputenc}
\usepackage[T1]{fontenc}
\usepackage{authblk}
\usepackage{graphicx}
\usepackage[]{lineno}
\usepackage{amsmath}
\usepackage{amssymb}
\usepackage[dvips]{geometry}
\title{$q$-Weibull distributions describing commercial service routes}

\author[1,*]{Ronan S. Ferreira}
\author[2]{Priscila C. A. da Silva}
\affil[1]{Universidade Federal de Ouro Preto, Departamento de Ciências Exatas e Aplicadas, João Monlevade, 35931-008, Brazil}
\affil[2]{Marinha do Brasil, Centro de Instrução Almirante Braz de Aguiar, Belém, 66.816-900, Brazil}
\affil[*]{ronan.ferreira@ufop.edu.br}
\date{}
\begin{document}
\maketitle
\begin{abstract}
We present an investigation of the mode of road transport in Brazil combining tools of complex networks and real-data. Our analysis involves a data-set based on the service routes inscribed on the Brazilian Transport Agency database. Although connectivity distributions of road networks worldwide are usually claimed as described by a power-law fashion, we report a better fit for the Brazilian case offered by the $q$-Weibull distribution. In our approach nodes assume the role of localities, whereas links represent service routes among them. Interestingly, a rapid drop takes place on the tail of the data distribution for a particular range of the number of outgoing connections. The mechanism responsible for driving this drop is revealed by investigating the spectral centrality of the network and different patterns of disassortative mixture, for both incoming and outgoing distributions. Besides a discussion about a power law description, we report a contrast with two different distributions. They are interpolated by the $q$-Weibull one: the Weibull and the $q$-exponential distributions. Moreover, we study the reciprocity of this network, which reflects the influence of mutual links over dynamical processes. This kind of analysis is indispensable for studies on human mobility, shipping, and a multi-modal perspective.
\end{abstract}
\vspace{0.25cm}
\noindent
{\bf keywords}: Networks; $q$-Weibull distributions; Complex Systems

\section{Introduction}
Structure and dynamics of systems arranged as directed networks composed of flows have attracted attention in many fields~\cite{xie2011evolving,bell1997transportation,batty2013new,marshall2004streets,acuto2018building}. Among studies of infrastructure systems, modes of transport have a particular interest since they are straightforwardly related to human mobility~\cite{cuttone2018understanding,barbosa2018human}, spreading disease \cite{strano2018mapping}, shipping~\cite{de2018distribution,gallardo2018energy}, \textit{etc}. Besides that, in recent years there have been many studies investigating structural features of transport systems through the application of complex network theory~\cite{dai2018evolving,ducruet2018maritime,zhu2018analysing,raimbault2018indirect,garcia2018analysis,mureddu2018complex,WU2018871}, which in turn offers materials for related works mainly on dynamical processes occurring on these structures.

Particularly, road transport is crucial for modern society in developing countries. Some of them have experienced a rapid growth in the human mobility for urban areas over last few decades by roads, such as in Brazil. According to the Brazilian Institute of Geography and Statistics (https://www.ibge.gov.br/) it is estimated the number of people living in urban areas is about 160 millions, corresponding to 84.4\% of the population (about 78 millions of individuals over last three decades). In 2017, about 88.7 million passengers were moved of which 78.9 million covered interstate and semi-urban transport, according to the Statistical Yearbook of Transport 2010-2017 (http://www.transportes.gov.br). To sum up, this modal is the most expressive in the country accounting for 485.625 Million TKU (Tonnes Useful Mile), corresponding to 61.1\% of all cargoes and passengers moved (the second largest modal of transportation is the railway with 164.809 millions TKU, representing 2.7\% per year \cite{cnt2018}).

In this work, our goal is to investigate aspects of the connectivity correlation and reciprocity of the Brazilian service routes concerning those companies registered on the National Land Transport Agency (ANTT) data-base (freely available at http://www.antt.gov.br/), acquired for the year of 2015: It is the starting point of a National Logistics Plan (http://www.epl.gov.br/) on the transportation modes, which aims at strategies for infrastructure optimization until 2035. Moreover, we are interested in tracing a mathematical description for incoming and outgoing connectivity distributions by revealing the arrangement of service routes among localities. Although we are going to talk about highways and imagine two-way roads, this makes sense when dealing with passenger transportation, but not necessarily in the case of goods exchange. This is a fundamental problem of several dynamical aspects and the measure of connectivity correlation and reciprocity are essential quantities concerning  systems composed of flows, arranged as directed networks~\cite{PhysRevLett.93.268701}. On one hand, the measure of reciprocity reflects the tendency for nodes in a network to be related by mutual links and this kind of analysis is fundamental for studies on patterns of human mobility and shipping. On the other hand, correlation is related with a measure of similarity of some features among (in our case) the number of incoming and outgoing connections among localities.

A complex network is a set of nodes interconnected by links according to some statistical distribution, for which we associate physical interpretations \cite{dorogovtsev2010}. In this paper $N=3240$ nodes assume the role of localities, whereas $L=104332$ links represent service routes. Three very questions are addressed: (\emph{i}) What are the entrance and exit patterns of the road network and the best-fitting description for the whole distribution arrangement? (\emph{ii}) How different are the set of incoming and outgoing service routes for each locality? (\emph{iii}) How reciprocal is this road network? 

\section{Methods}
Information about connections between localities are freely available in the form of a list of links (http://www.antt.gov.br/). In the mathematical language, this list is an array of kind $\mathbf{E}:L\times2$, where $L$ is the total number of roads or service routes connecting localities by pairs. Therefore, this array has $L$ lines and $2$ columns: In the first column are placed localities of origin and in the second one destinations (an OD-matrix). The elements $\{e_{i,j}\}$ of $\mathbf{E}$ follow the following rule: $e_{i,j=1}=m$ (where $m$ is the number labeling the origin locality) and $e_{i,j=2}=n$ (where
$n$ represents the label of the destination locality) if there is a service route from $m$ to $n$, which in turn is registered at least one company in the ANTT data-set. So, the matrix $\mathbf{E}$ maps flow of directions between localities in the network. Note that is also possible to know the number of different companies that travel on each route, revealing a character of density of companies for each of them.

We can construct a methodology by means of mapping $\mathbf{E}$ onto an adjacency matrix $\mathbf{A}:N\times N$, where
$N$ is the total number of nodes. The elements $\{a_{ij}\}$ of $\mathbf{A}$ are set to unit if there is a connection not \emph{between} $i$ and $j$, but \emph{from} $i$ to $j$ ($a_{ij}=0$, otherwise). Conversely, $a_{ji}=1$ if the flow also goes in
opposite direction, that is, from $j$ to $i$. The indices $i$ and $j$ correspond to the numbers $m$ and $n$ that are read straightforwardly from the matrix $\mathbf{E}$, columns $1$ and $2$, respectively. By means of this adjacency matrix is possible to study the mode of road transport using complex network theory. For example, the number of outgoing service routes that a node $i$ has with others can be obtained by summing over his adjacency matrix line. Equivalently, summing over the column of the node $i$, one can compute its number of incoming service routes. So, the road network is directed, which in turn means that links indicate a flow direction between localities. For this reason, nodes have two different degrees: (\emph{i}) \emph{in}-degree, the number of incoming links and (\emph{ii}) \emph{out}-degree, the number of those outgoing connections. One may obtain, these numbers computing
\begin{eqnarray}
  k_{i}^{out}&=&\sum_{j=1}^{N}a_{ij}\label{eq:kout}\\
  k_{i}^{in}&=&\sum_{j=1}^{N}a_{ji}\label{eq:kin},
\end{eqnarray}
where $k^{out}$ counts the \emph{out}-degree, whereas $k^{in}$ the \emph{in}-degree, respectively for equation (\ref{eq:kout}) and (\ref{eq:kin}). In general terms, letting the number of nodes of degree $k_{\beta}$ be $N_{k_{\beta}}$ (where $\beta$ denotes \emph{in} or \emph{out}), $\sum_{k_{\beta}}N_{k_{\beta}}=N_{\beta}$, we can estimate the frequency of occurrence, $P(k_{\beta})$, of a node with $k_{\beta}$ connections. Equivalently, we can write $P(k_{\beta})$ as the fraction of nodes of $k_{\beta}$ connections in the network,
\begin{equation}
P(k_{\beta})=\frac{N_{k_{\beta}}}{N_{\beta}}.
\label{eq03}
\end{equation}
From these distributions concerning on incoming and outgoing connections, we can drawn and analyze the road network profile and perform analysis on correlations and reciprocity on the network.

\section{Results}
\label{sec:res}
\subsection{Incoming and outgoing connections profile}
\begin{figure}[h]
\centering
\includegraphics[scale=0.35]{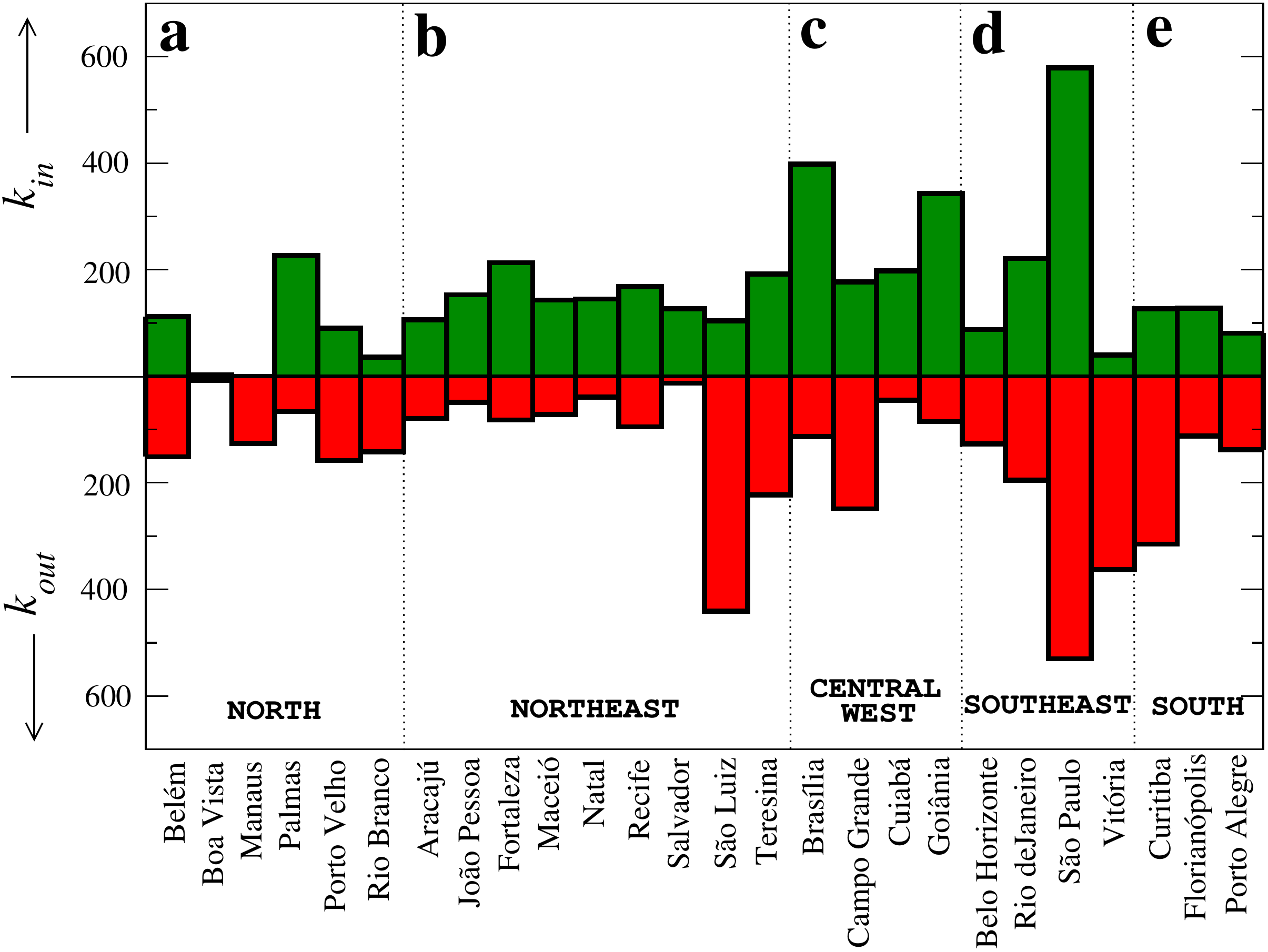}
\caption{
\label{fig:3} \textbf{Balance of service routes of Brazilian capitals.} Number $k_{in}$ of incoming (arrow up) and $k_{out}$ of outgoing (arrow down) service routes for each one of the $26$ Brazilian capitals. The Brazilian federalization is divided by five regions: $\mathbf{a}$ North, $\mathbf{b}$ Northeast, $\mathbf{c}$ Central-West, $\mathbf{d}$ Southeast, and $\mathbf{e}$ South. To trace the profile of the entire road network, we used equation (\ref{eq03}) to obtain the histograms of frequency of occurrence as a function of the number of connections of each locality.}
\end{figure}

\begin{figure}[h]
\centering
\includegraphics[scale=0.32]{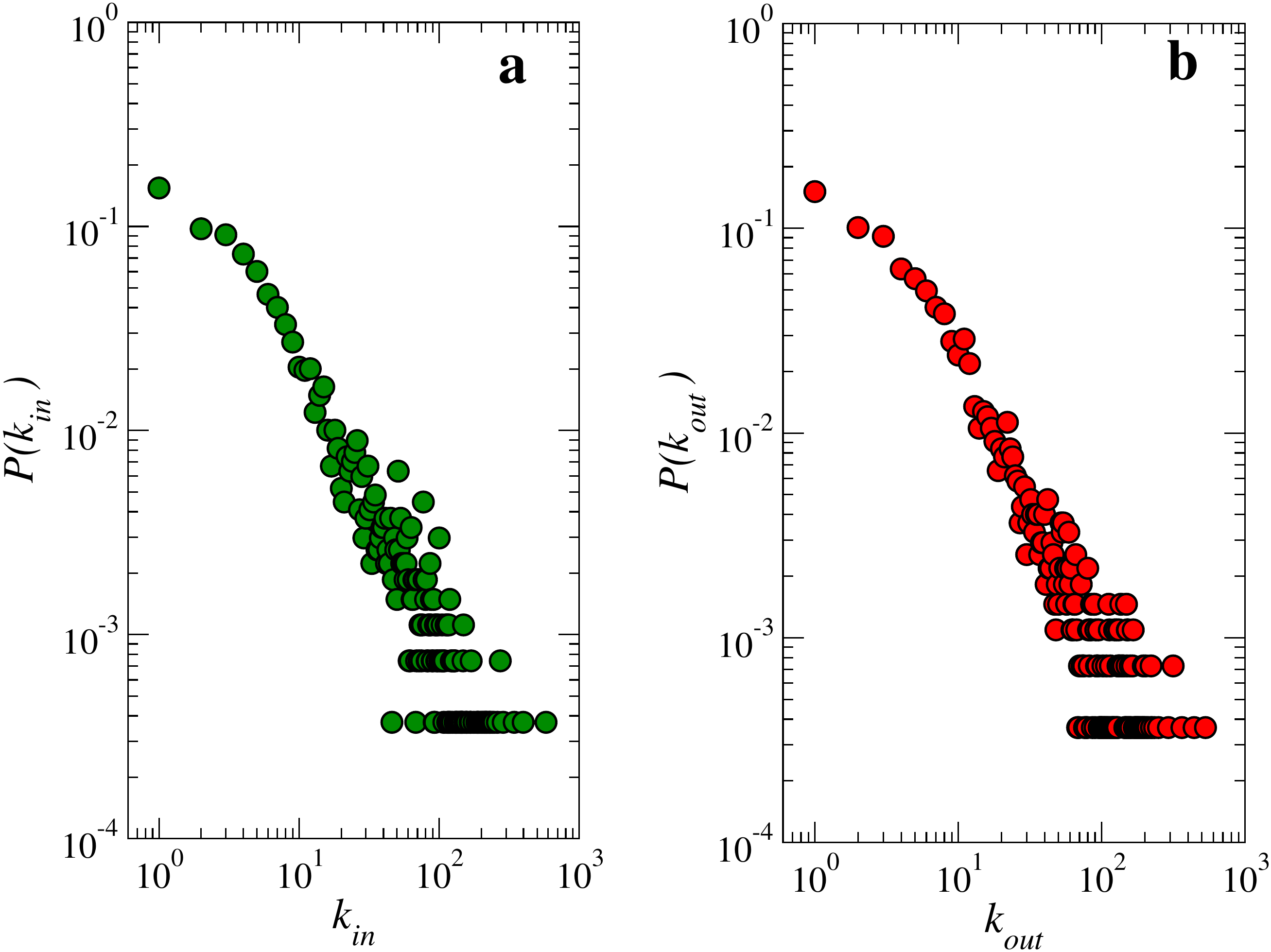}
\caption{\label{fig:2} \textbf{The complete profile of connections.} Histogram $P(k_{\beta})$ vs. $k_\beta$ of discrete probabilities for the number of incoming routes $\beta\equiv$~$in$ in frame $\mathbf{a}$ and the number of outgoing connections $\beta\equiv$~$out$ in frame $\mathbf{b}$ for companies cataloged in the Brazilian National Land Transport Agency database. Both frames show statistical fluctuations at the tail of the distributions for which we performed their cumulative versions (next Figure) using equation (\ref{eq:cum}).}
\end{figure}

We compile for the sake of comparison and to introduce the subject of connectivity (degree) distribution the degree $k$ (number of connections) for the twenty six Brazilian capitals. Figure \ref{fig:3} shows the output corresponding to \emph{in}- and \emph{out}-degree for each capital, categorized by the five regions in which Brazilian's federalization is divided: ($\mathbf{a}$) North region with 6 capitals and 14\% of service routes, followed by ($\mathbf{b}$) Northeast, with 9 capitals responsible for 30\% of service routes, and ($\mathbf{c}$) Central-West, with 4 capitals and 19\% of service routes. Next, ($\mathbf{d}$) Southeast also with 4 capitals and 26\% of the total service routes and, finally, ($\mathbf{e}$) South region with 3 capitals, responsible for 11\% of the total of service routes among capitals acquired from ANTT database. It is a set of regional centers which in turn is not necessarily interconnected, although important to the characterization of the profile of the entire connectivity distribution. It is possible to recognize them located in the tail of curves presented in Figure \ref{fig:2}, identifying them as \emph{hubs} from a network point of view.

Carrying on the number of incoming and outgoing connections of all localities in the network, respectively denoted by $k_{in}$ and $k_{out}$ and the definition provided from equation (\ref{eq03}), we obtained the full connectivity distribution of service routes. In terms of \emph{in}- and \emph{out}-degree, one can write down equations (\ref{equation Pkin}) and (\ref{equation Pkout}) from equation (\ref{eq03}). Since $\sum N_{k_{in}}=N_{in}$ and $\sum N_{k_{out}}=N_{out}$, for $k_{in}>0$ and $k_{out}>0$, we identified equations (\ref{equation Pkin}) and (\ref{equation Pkout}) as normalized histograms of discrete probabilities of finding a locality of degree $k_{in}$ and $k_{out}$:
\begin{equation}
P(k_{in})=\frac{N_{k_{in}}}{N_{in}},
\label{equation Pkin}
\end{equation}
\begin{equation}
P(k_{out})=\frac{N_{k_{out}}}{N_{out}}.
\label{equation Pkout}
\end{equation}
The results obtained from these analysis are shown in Figure \ref{fig:2}-$\mathbf{a}$ for the \emph{in}-degree distribution and \ref{fig:2}-$\mathbf{b}$ for the \emph{out}-degree distribution. Both curves show at first glance the same logarithmic functional behavior on log-log scales, revealing a heterogeneous feature of these distributions. The highest values of $k_{in}$ shown in Figure \ref{fig:3} are located on the tail of the distribution displayed in Figure \ref{fig:2}-$\mathbf{a}$, whereas the lowest ones are shown in the head of the \emph{in}-degree distribution (same frame). The same analysis can be performed for the highest and the lowest values of $k_{out}$ in the \emph{out}-degree distribution exhibited in Figure \ref{fig:2}-$\mathbf{b}$. Those with high values of $k$ are harder to find in the network, corresponding on an occurrence of order of one part in $10^4$. They are related to the urban and regional centers in the country.

We also treated statistical fluctuations of the data presented at the tail of the degree distributions by computing their cumulative versions. We assume $f(x)=\int^{\infty}_xg(x')dx'$, where $x$ is a random variable corresponding to the also random variable $x'$, truncated inside $[x,\infty)$. In our case, dealing with discrete variables we denote this new version of the original distribution $P(k_{\beta})$, by $p(k_{\beta})$, where the indice $\beta$ can denotes \emph{in} or \emph{out} tags:
\begin{equation}
p(k_{\beta})=
\sum_{k_{\beta}'\geq k_{\beta}}P(k_{\beta}').
\label{eq:cum}
\end{equation}

\begin{figure}[h]
\centering
\includegraphics[scale=0.32]{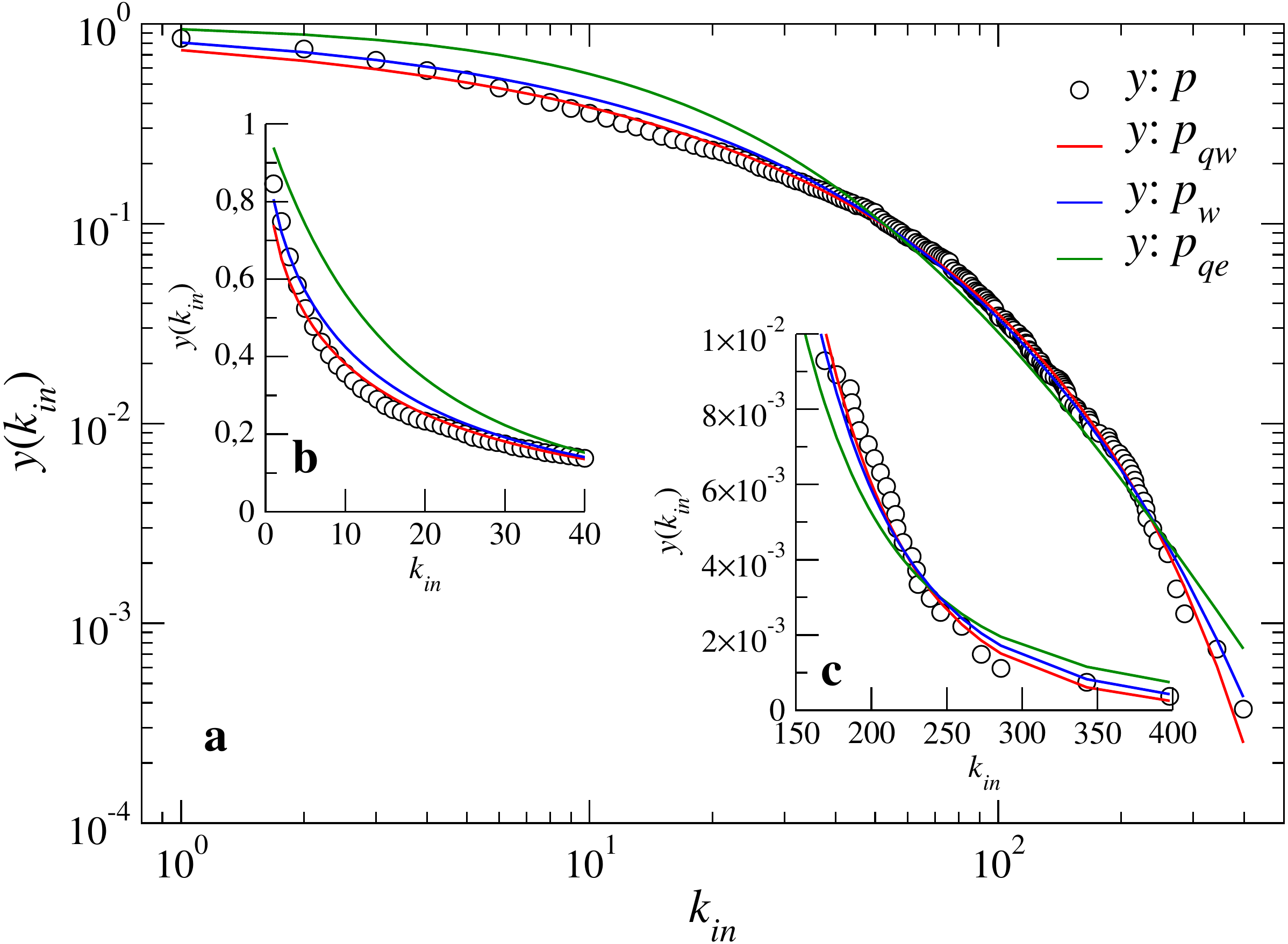}~\includegraphics[scale=0.32]{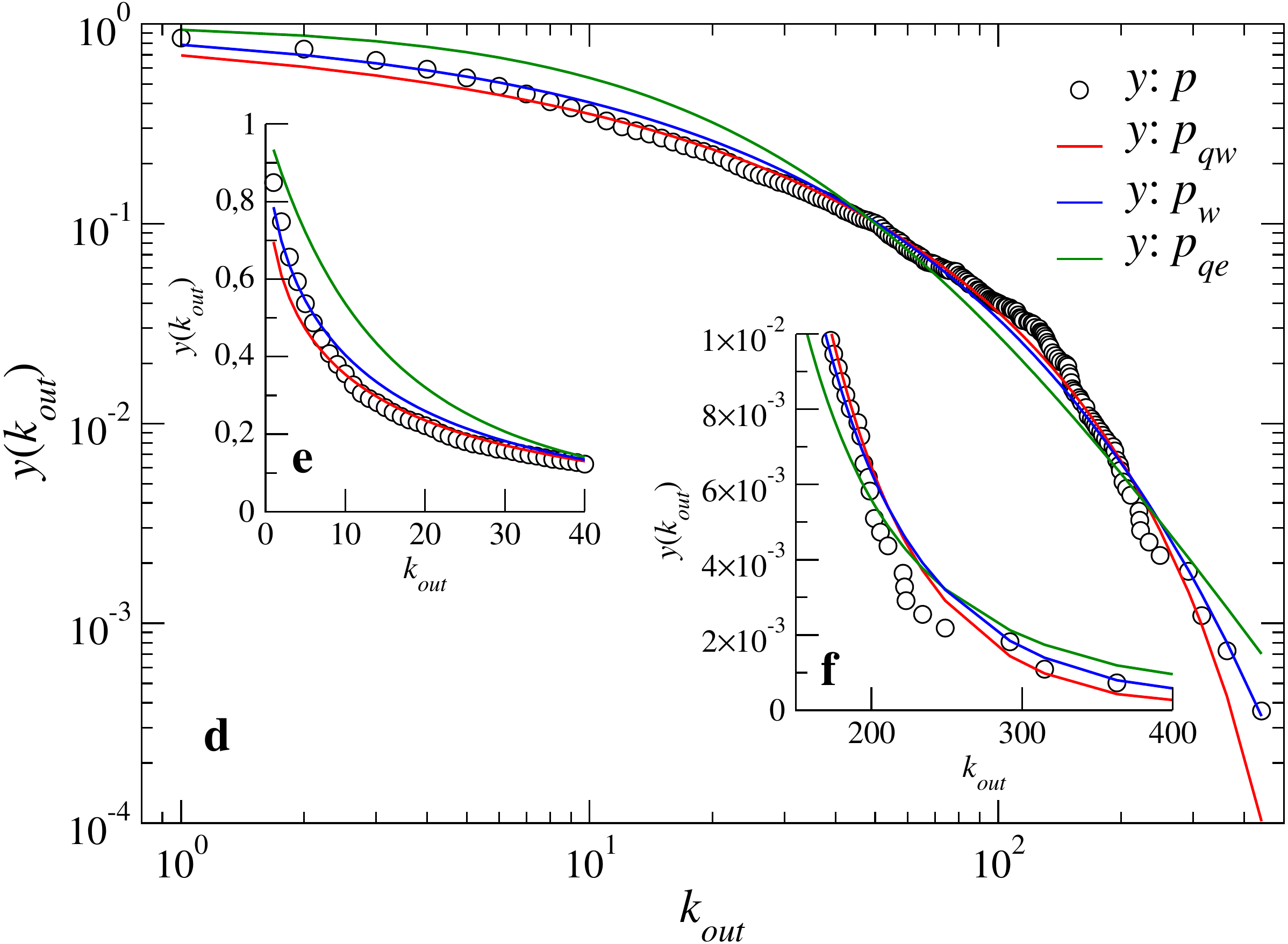}
\caption{\label{fig:acum} \textbf{Analysis of the cumulative versions.} Cumulative versions (circles) of distributions shown in previous figure. Lines are adjusts following equations (\ref{eq:pqw}), (\ref{eq:pw}), and (\ref{eq:pqe}). The best-fitting shown in $\mathbf{a}$ was obtained for the incoming connections of routes using $q$-Weibull distribution with the parameters reported on the third column of Table \ref{tab:par}. Inset $\mathbf{b}$ shows adjusts at the head of the \emph{in}-distribution, whereas inset $\mathbf{c}$ at its tail. The same organization of frame $\mathbf{d}$ and insets $\mathbf{e}$ and $\mathbf{f}$ is shown for \emph{out}-degree distribution. In the range of $k_{out}$ about the order of $10^2$ appear a faster drop when compared to the same interval for $k_{in}$. The mechanism responsible for this is studied below.}
\end{figure}

The outputs from analysis performed from equation (\ref{eq:cum}) are shown by means of symbols (circles) in Figure \ref{fig:acum}. In the frame \ref{fig:acum}-$\mathbf{a}$ is presented the cumulative $in$-degree distribution $p(k_{in})$ and frames \ref{fig:acum}-$\mathbf{b}$ and \ref{fig:acum}-$\mathbf{c}$ show in more details the head and the tail of this distribution, respectively. The same organization (main plot and insets) for the \emph{out}-degree distribution $p(k_{out})$ is shown on frames \ref{fig:acum}-$\mathbf{d}$, \ref{fig:acum}-$\mathbf{e}$, and \ref{fig:acum}-$\mathbf{f}$. Unlike these decays in frames \ref{fig:acum}-$\mathbf{a}$ and \ref{fig:acum}-$\mathbf{c}$ their non-cumulative versions on Figure \ref{fig:2} have a typical power-law fashion as reported in Ref.~\cite{Kalapala2006ScaleII}, although in that reference the authors investigated road networks in United States, England and Denmark, taking into account their geographical structure from their dual representation. In that approach roads acts as nodes and a link connects two nodes, if the corresponding road ever intersect. Nevertheless, the power-law description of these distributions should be observed even on their cumulative versions, $P(x)=Cx^\alpha\implies p(x)=C'\left[x'^{\alpha+1}\right]_x^\infty$. Since the cumulative version of a power-law also exhibits a power-law fashion, decaying as a linear curve on a \emph{log-log} scale as set on Figure \ref{fig:acum}. Moreover, the power-law hypothesis we have studied fails under the Likelihood test \cite{clauset2009power} - data not shown. However, in this section we give attention to others distributions, namely the $q$-Weibull, the Weibull, and the $q$-exponential distributions. In the section \ref{sec:pl} we return to the subject of the fail of the power-law description for the data-set we have used.

For this data-set what we observed was a steep decay for large values of degree $k$, kind of investigated in Ref.~\cite{picoli2003q}. Following this reference, we studied the curves we found using $q$-distributions~\cite{picoli2009q}. Particularly, we found a good fit by using the $q$-Weibull one, which is given by the expression
\begin{equation}
P_{qw}(x)=p_0\frac{rx^{r-1}}{x_0^r} \exp_q\left[-\left(\frac{x}{x_0}\right)^r\right],
\label{eq:Pqw}
\end{equation}
where $exp_q(-x)\equiv\left[1-(1-q)x\right]^{1/(1-q)}$ if $\left[1-(1-q)x\right]\geq 0$ and $exp_q(-x)\equiv 0$, otherwise~\cite{tsallis2009introduction}. Computing the cumulative version of equation (\ref{eq:Pqw}) one obtains 
\begin{equation}
p_{qw}(x)=p'_{0}\exp'_{q}\left[-\left(\frac{x}{x'_{0}}\right)^r\right],
\label{eq:pqw}
\end{equation}
with $q'=1/(2-q)$, $x'_{0}=x_{0}/(2-q)^{1/r}$ e $p'_{0}=p_{0}/(2-q)$. This is shown in Figure~\ref{fig:acum} besides the outputs reached by performing adjusts for the cumulative versions (circles) of the data presented in Figure~\ref{fig:2}. The best-fitting of curves were achieved using equation~(\ref{eq:pqw}) - guided by the statistical test of the relative Root Mean Square error (RMSe), with the parameters (non weighted) reported on Table~\ref{tab:par}. For a comparison, we let in this Figure~\ref{fig:acum} fits reached by two different distributions. The $q$-Weibull interpolates between them: The (pure) Weibull distribution and the $q$-exponential one. In the limit of $q\rightarrow1$, equation (\ref{eq:Pqw}) recovers the Weibull distribution, which is given by $P_w(x) = p_0\frac{rx^{r-1}}{x_0^r}\exp{\left[-\left(\frac{x}{x_0}\right)^r\right]}$ with its cumulative version,
\begin{equation}
    p_{w}(x) = p_0\exp{\left[-\left(\frac{x}{x_0}\right)^r\right]}.
\label{eq:pw}    
\end{equation}
On the other hand, in the limit of $r\rightarrow1$, equation (\ref{eq:Pqw}) recover the $q$-exponential distribution, $P_{qe}(x)=p_0\exp_q{\left[-\left(\frac{x}{x_0}\right)^r\right]}$ for $\left[1-(1-q)(x/x_0)\right]\ge0$ and normalized if $p_0=(2-q)/x_0$. Observe that in the limit $q\rightarrow1$, the $q$-exponential recovers the exponential distribution. Its cumulative version is given by
\begin{equation}
p_{qe}(x)=p'_0\exp'_q{\left[-\left(\frac{x}{x_0}\right)^r\right]}
\label{eq:pqe}
\end{equation}
for $q<2$ with $q'=1/(2-q)$, $x'_0=x_0/(2-q)$, and $p'_0=p_0x_0/(2-q)$.

Both \emph{in}- and \emph{out}-distribution were better-fitted by the equation (\ref{eq:pqw}). However, for values of $k$ greater than the order of $10^2$ on the outgoing distribution - frame \ref{fig:acum}-$\mathbf{d}$, we observed a faster drop, compared with the same range of values of $k_{in}$. The mechanism responsible for this will be presented in the section \ref{sec:a} and it is related to different patterns of degree mixture for \emph{in}- and \emph{out}-distribution.
\begin{table}[h]
\centering
\begin{small}
\begin{tabular}{cccccc}
\hline
 & & \multicolumn{2}{c}{non weighted} & \multicolumn{2}{c}{weighted} \\
 \hline
& Parameter & $p(k_{in})$ & $p(k_{out})$ &
 $p(s_{in})$ & $p(s_{out})$\\
\hline
         & $q$    & $ 0.883$ & $ 0.824$ & $ 0.730$ & $ 0.844$\\
$p_{qw}$ & $r$    & $ 0.489$ & $ 0.435$ & $ 0.332$ & $ 0.363$\\
         & $x_0$  & $15.025$ & $15.909$ & $34.813$ & $22.155$\\
         & \tiny{RMSe}  & $0.078$  & $0.145$ & $0.099$   & $0.079$\\
\\
$p_{w}$  & $r$    & $0.598$  & $ 0.575$ & $ 0.512$ & $ 0.467$\\
         & $x_0$  & $12.977$ & $11.884$ & $20.430$ & $16.435$\\
         & \tiny{RMSe}  & $0.107$ & $0.173$ & $0.178$ & $0.111$\\
\\
         & $q$    & $ 1.229$  & $ 1.249$  & $ 1.296$ & $ 1.334$\\
$p_{qe}$ & $x_0$  & $12.402$  & $10.877$  & $17.683$ & $14.189$\\
         & \tiny{RMSe}  & $0.167$   & $0.316$   & $0.374$ & $0.317$\\
\hline
\end{tabular}
\caption{
\label{tab:par} \textbf{Table of parameters.} The parameters used for the adjusts present on Figure \ref{fig:acum} given by equations (\ref{eq:pqw}), (\ref{eq:pw}), and (\ref{eq:pqe}) are present in the third and fourth columns. At fifth and sixth columns, the parameters are related to adjusts shown on Figure \ref{fig:Sqw}, where were used weighted cumulative versions (discussed below) of those degree distributions present on Figure {\ref{fig:2}}.}
\end{small}
\end{table}

\subsection{Analysis of the eigenvalue centrality}
\label{sec:a}
\begin{figure}[h]
\centering
\includegraphics[scale=0.35]{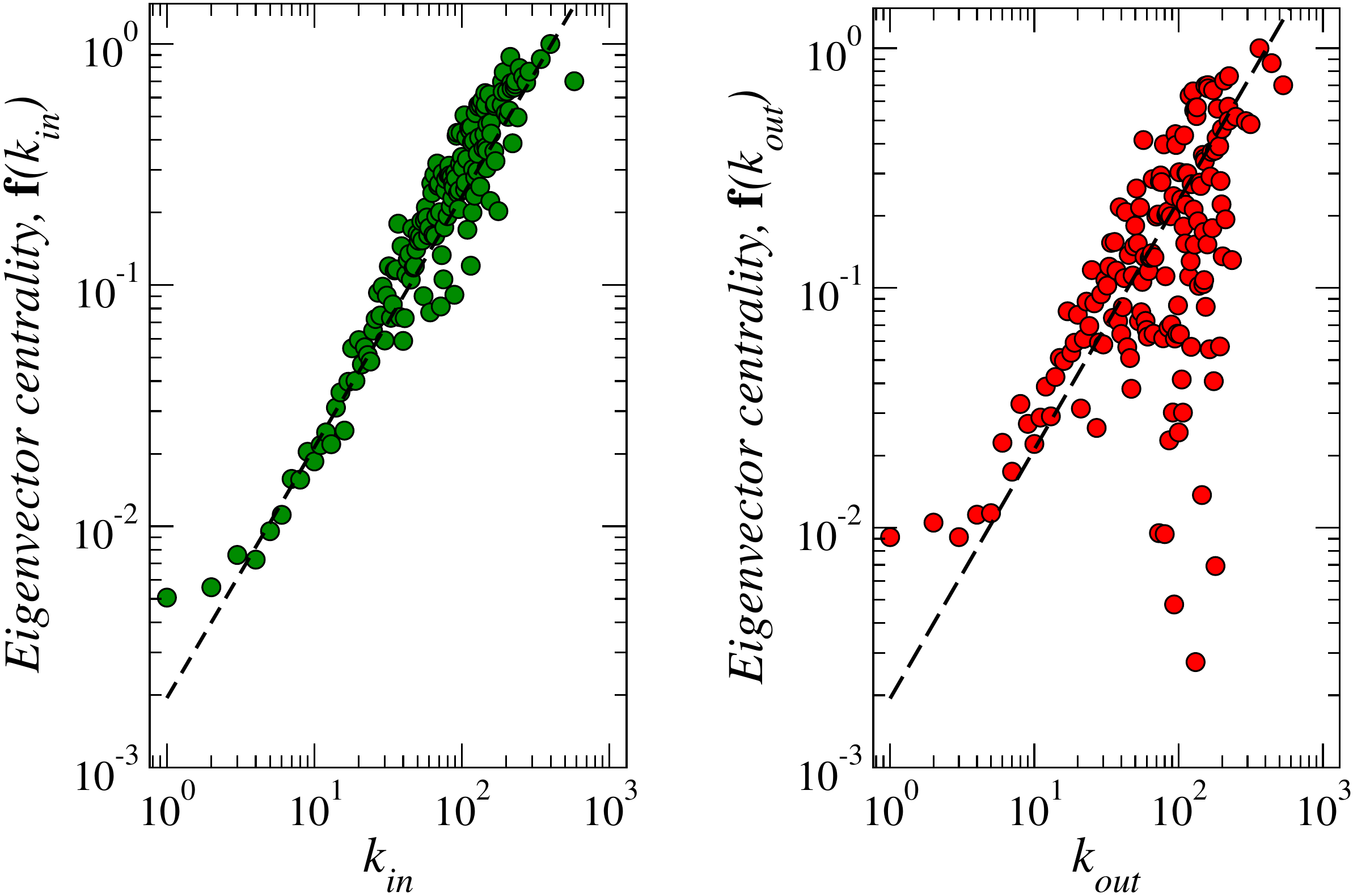}
\caption{ \label{fig:bc} \textbf{Eigenvector centrality.}  $\mathbf{f}(k_{\beta})$ as a function of the number of incoming routes $\beta\equiv$~$in$ on the left and the number of outgoing connections $\beta\equiv$~$out$ on the right. It is proportional to the number of random walks reaching each locality and reveals the importance of each of them based on the traffic over the network (Dashed lines are simple $y\propto x$ curves). A lack of this proportionality is observed pronounced in the range $60<k_{out}<200$ and is related to the steep decay shown in frame $\mathbf{d}$ in Figure \ref{fig:acum}.}
\end{figure}

The mechanism responsible for the steep decay mentioned in the previous section was revealed by performing a spectral analyze for the so-called eigenvector centrality \cite{newman2016mathematics}. It acknowledges that links are not equal and they have different influences on the network. In general, connections to urban centers will lend another locality more influence than connections to less influential places~\cite{bonacich1972technique}. We can allow for this effect by defining the centrality $\mathit{f}_i$ of a node $i$ as proportional to the average of this quantity inside its neighborhood
\begin{equation}
\mathit{f}_i=
\frac{1}{\lambda}\sum_{j=1}^Na_{ij}\mathit{f}_i,
\end{equation}
which in turn, on a matrix notation, yields $\mathbf{A}\mathbf{f}=\lambda \mathbf{f}$, with $\lambda\neq 0$ being a constant value. It is equivalent to assuming that a node is important if it is linked to other important nodes. Since $\lambda$ is the leading eigenvalue of $\mathbf{A}$, $\mathbf{f}$ can be obtained by performing the power method,
\begin{equation}
\mathbf{f}^{(n)}=\mathbf{f}^{(n-1)}\mathbf{A}
\end{equation}
using positive entries. Choosing $\mathbf{f}^{(0)}=(1,1,...1)$, $\mathbf{f}^{(1)}$ will correspond to the degree $k$ of each node and $n\ge1$ coincides to the number of walks of length $n$ between each pairs of nodes. Therefore, central nodes are
more visited than peripherical ones, proportional to the number of visits through a random walk of infinite length. Since measures of centrality can be defined in many ways, we choose this one due the transport character of the data-set we used. Figure \ref{fig:bc} shows a drop on the measuring of eigenvector centrality for those nodes with values of $k_{out}$ of the order of $10^2$ in discrepancy with the eigenvector analysis for the data-set concerning on values of $k_{in}$.

\begin{figure}[h]
\centering
\includegraphics[scale=0.25]{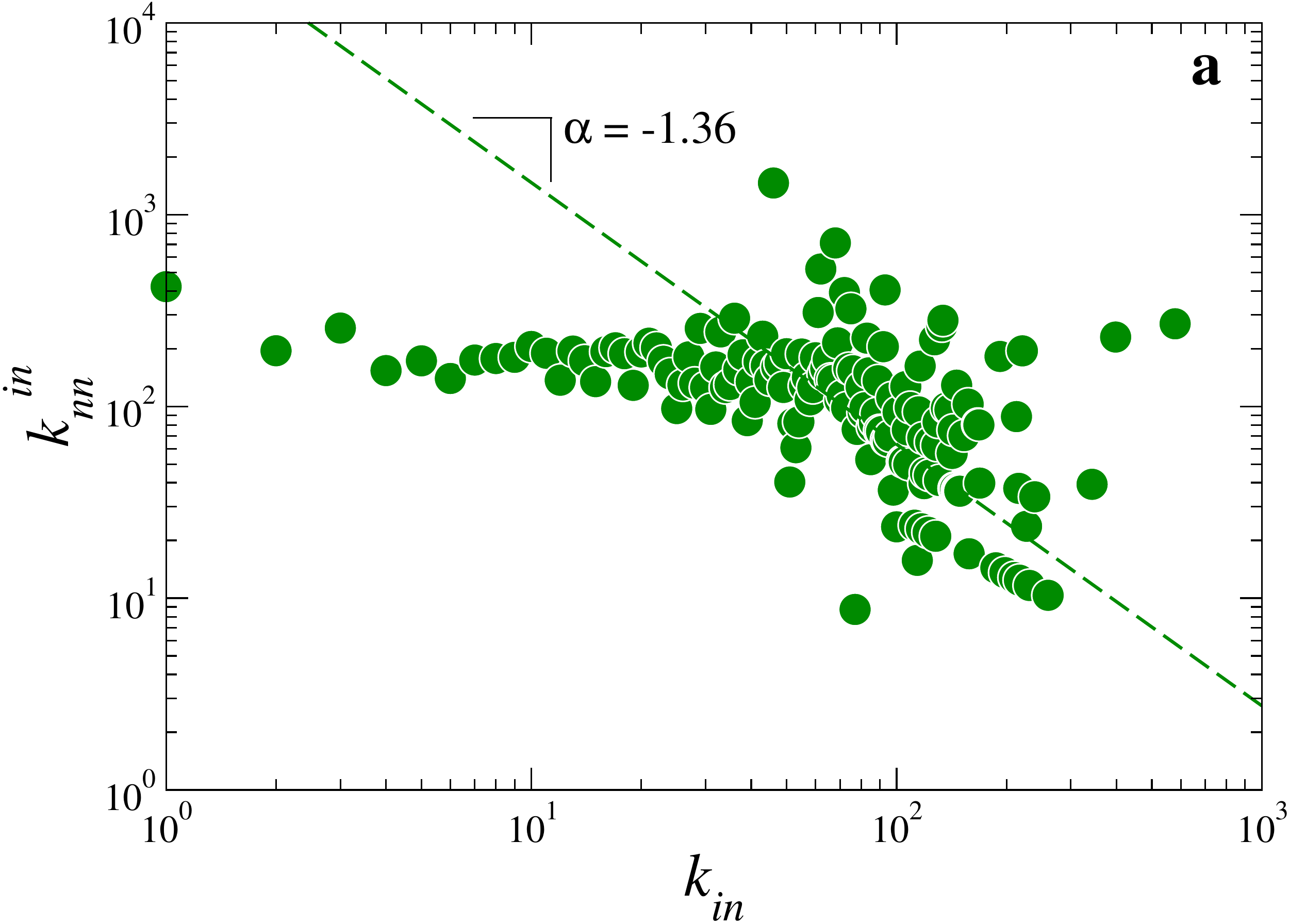}\includegraphics[scale=0.25]{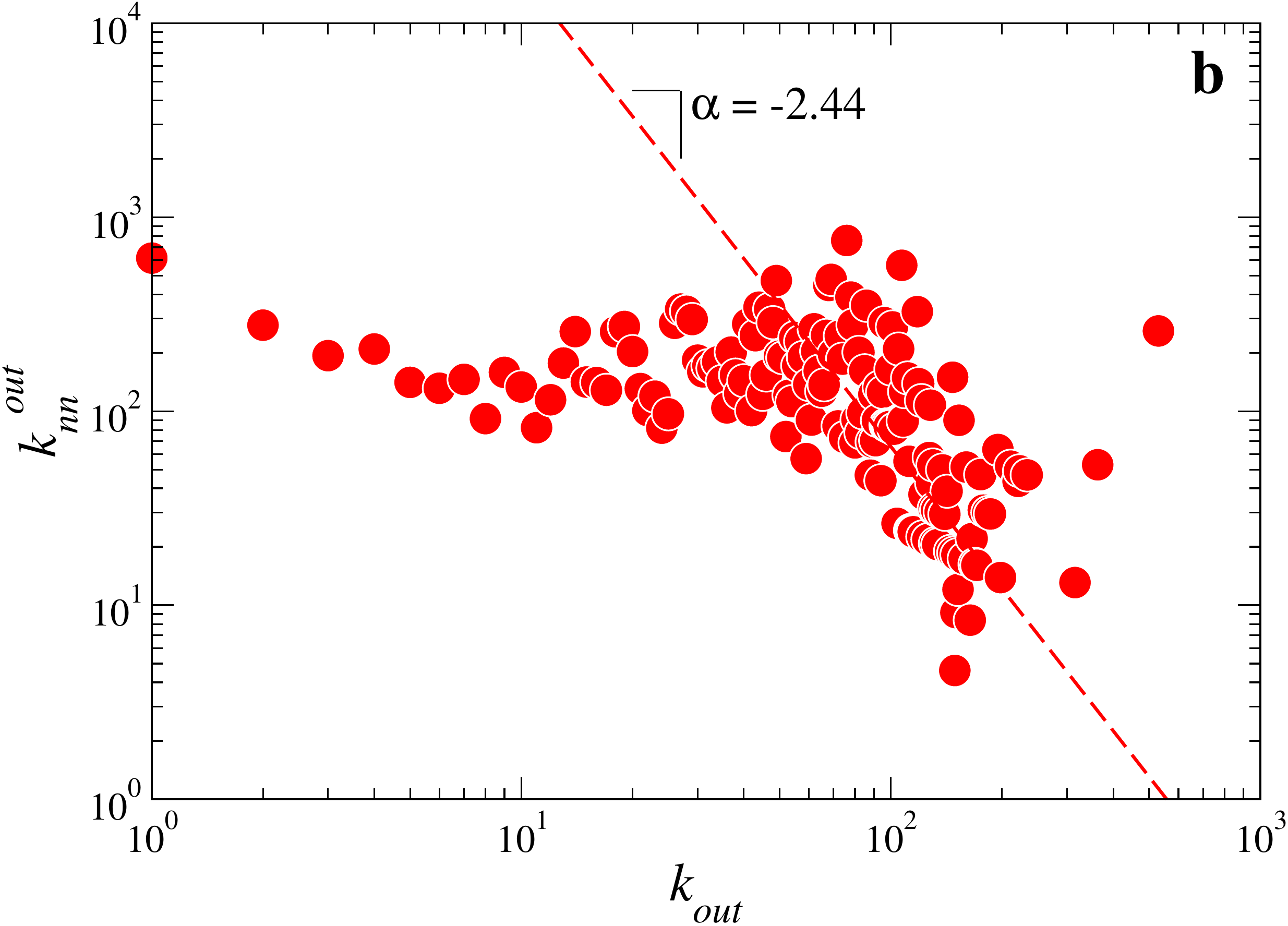}\\
\includegraphics[scale=0.25]{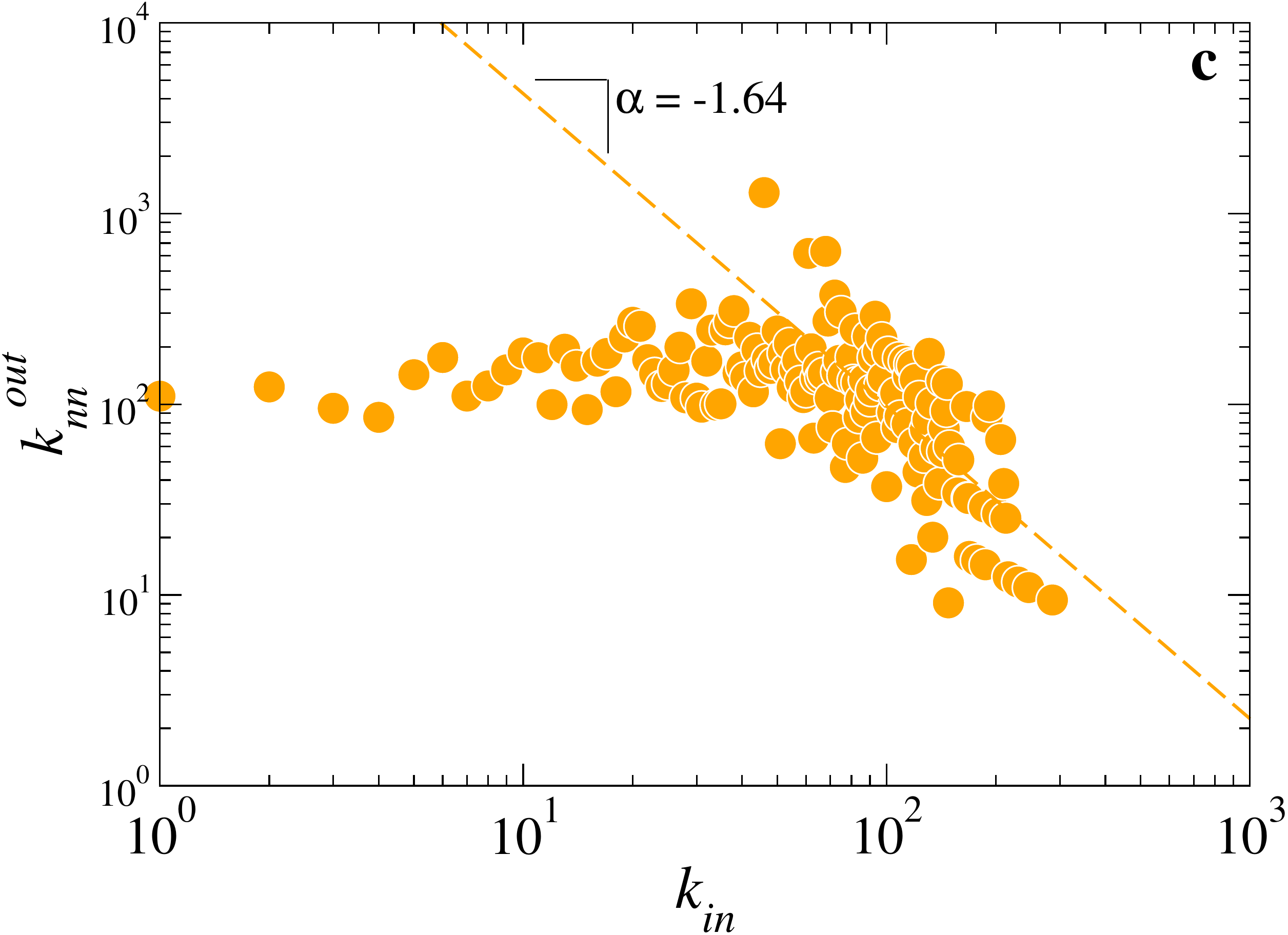}\includegraphics[scale=0.25]{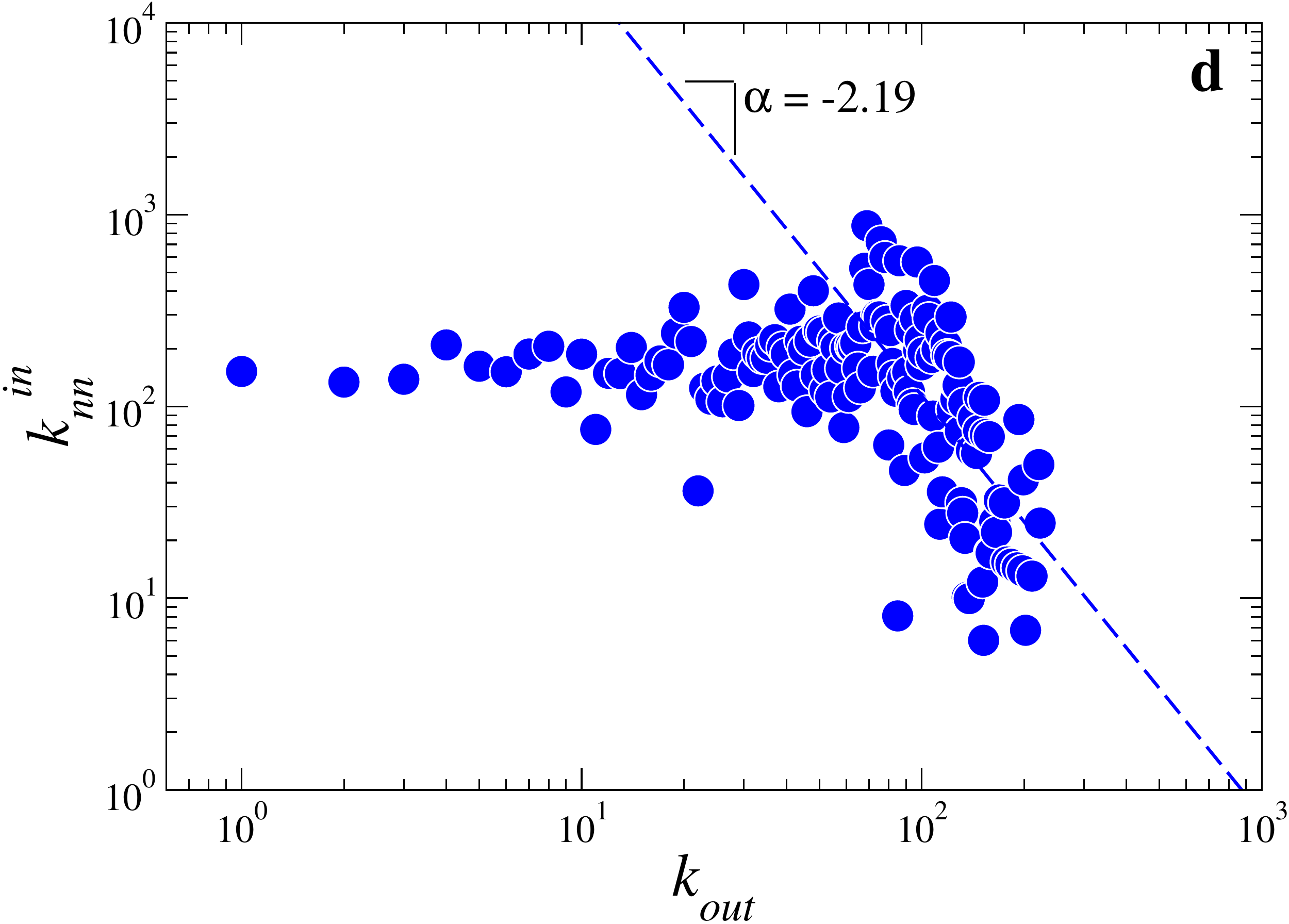}
\caption{\label{fig:knn} \textbf{Degree mixture.} The average nearest neighbor degree $k_{nn}^{\beta}$ as a function of $k_{in}$ ($\mathbf{a}$) and $k_{out}$ ($\mathbf{b}$). In frames $\mathbf{c}$ and $\mathbf{d}$ are shown $k_{nn}^{out}$ against $k_{in}$ and $k_{nn}^{in}$ against $k_{out}$, respectively. For the range $60<k_{in},k_{out}<200$ a disassortative degree mixture take place, in which nodes inside this range of degree are linked to nodes of smaller values of $k_\beta$, where $\beta$ depends for each one of these four cases. Assuming for that interval a functional relationship following a power-law fashion, these decays are marked (dashed lines) by $k_{nn}^{\beta}\propto k_{\beta}^{\alpha}$. The more pronounced decay for outgoing connections in frame $\mathbf{b}$ is related to the steep decay shown in Figure \ref{fig:acum}. The discussion concerning on frames $\mathbf{c}$ and $\mathbf{d}$ are presented in the end of section \ref{sec:res}.}
\end{figure}

The pronounced non proportionality present in the range $60<k_{out}<200$ is due to a strongest disassortative character present in the degree mixture arrangement found out in the neighborhood of nodes of degree in this range. In this scenario urban or regional centers avoid each other, linking instead to low-degree localities. It is in accordance if we think about the structure for a ``chain supply'' in a country with continental dimensions like Brazil (48\% of the total area of the South America). Figure \ref{fig:knn} shows some analysis for the average degree of the nearest neighbors, $k_{nn}$, as a function of the degree $k$ given by the expression~\cite{newman2018networks} $k_{nn}(k)=\sum_{k'}k'P(k'|k)$, where $P(k'|k)$ is the conditional probability that following a link of a node of degree $k$ we found at the other end a node of degree $k'$. It was used a generalization $k_{nn}^{in,out}$ $vs.$ $k_{in,out}$ to properly investigate the data-set present in Figure \ref{fig:bc}. We observed in Figure \ref{fig:knn}-$\mathbf{a}$ and \ref{fig:knn}-$\mathbf{b}$ a disassortative mixture for both cases, according to a power-law fashion adjusting on the range discussed above. Although both distributions are marked by a disassortative mixture in that range, a more clear-cut drop was noticed for the outgoing distribution ($k_{nn}^{in}\propto k_{in}^{-1.36}$ and $k_{nn}^{out}\propto k_{out}^{-2.44}$).

\subsection{Assigning weight to the connections}
\begin{figure}[h]
\centering
\includegraphics[scale=0.32]{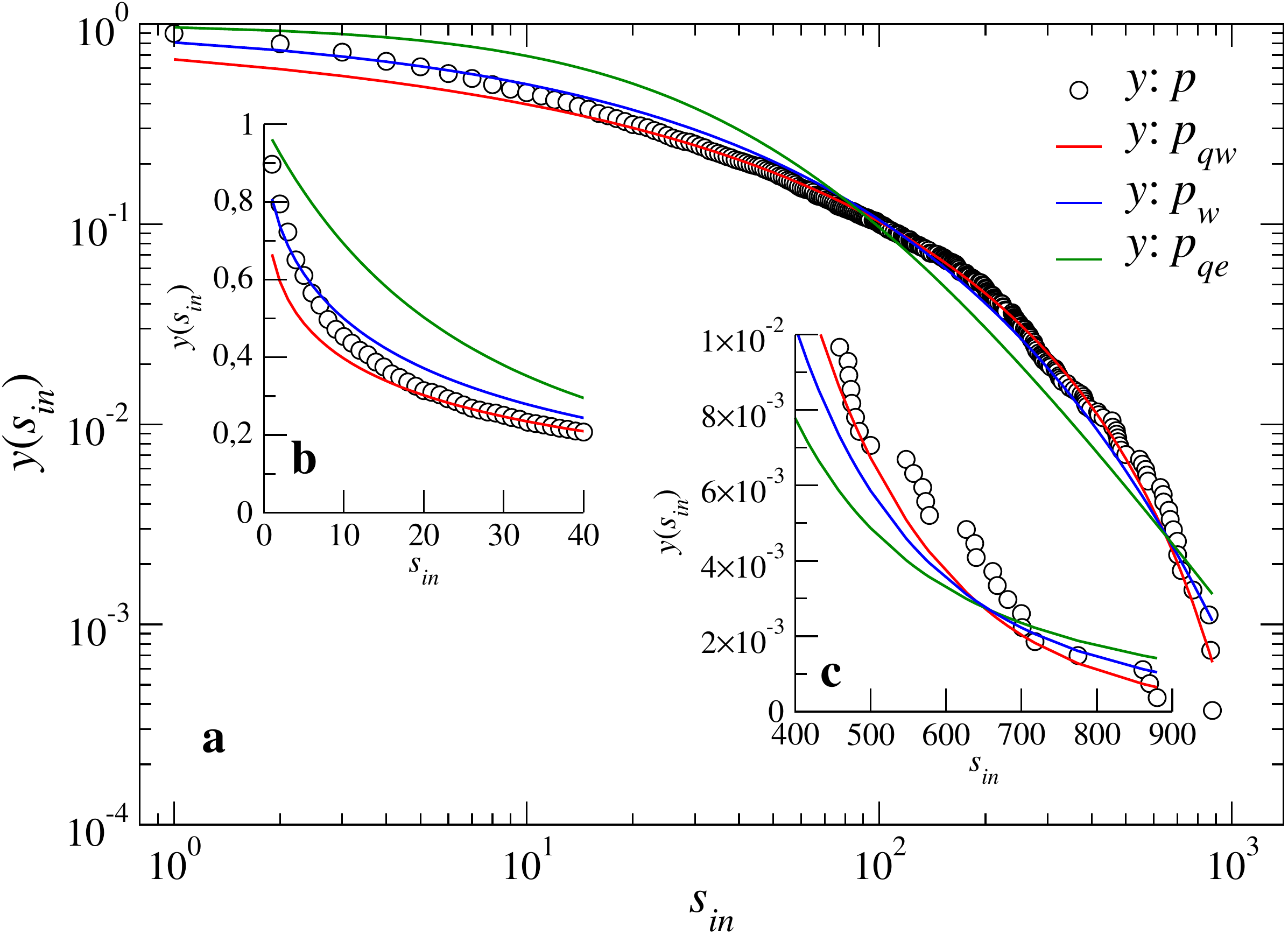}~\includegraphics[scale=0.32]{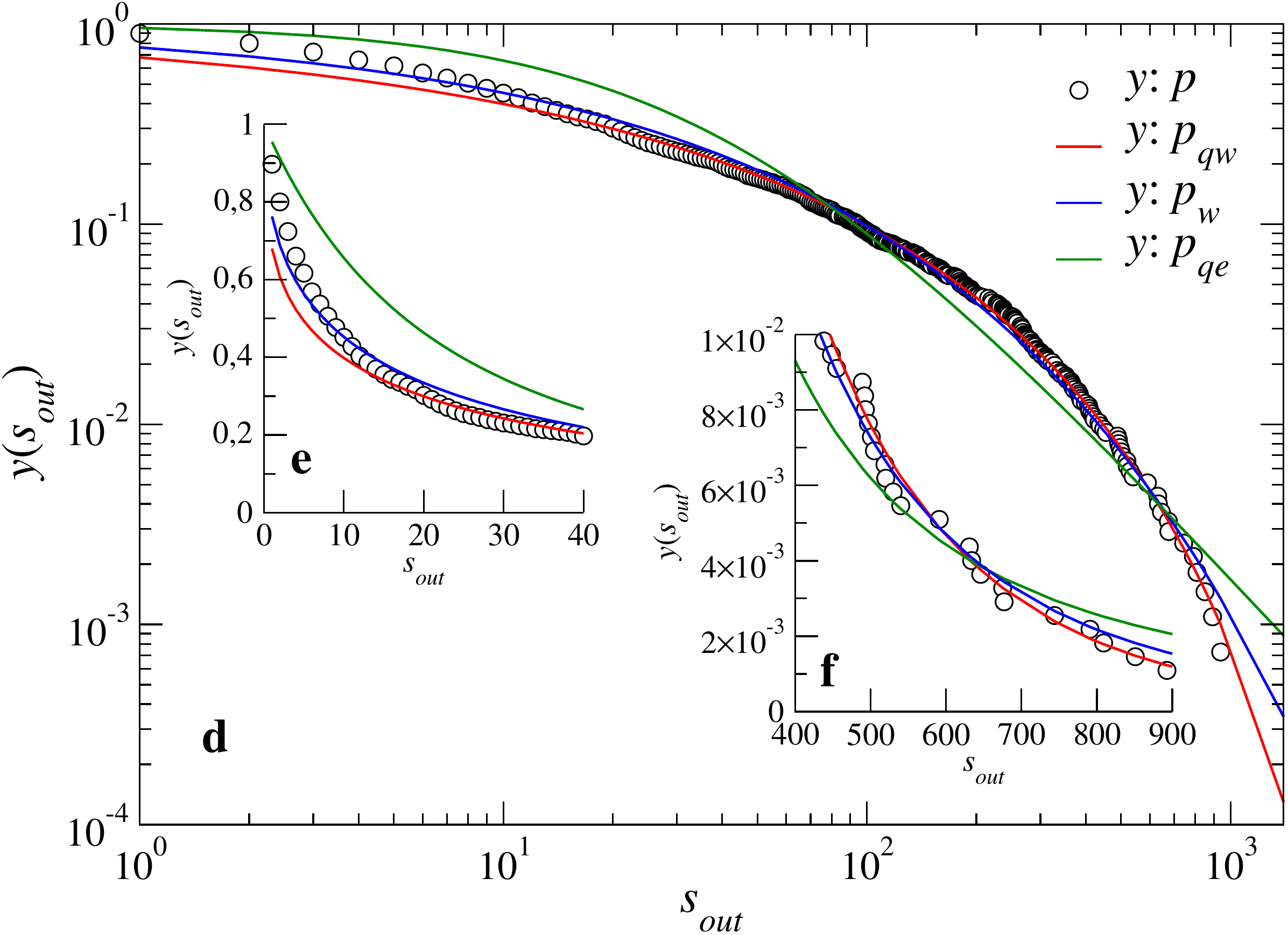}
\caption{\label{fig:Sqw} \textbf{Weighted network.} Service routes are weighted according to the density of companies, $w_{ij}$, that travel in them. It is a weighted cumulative version for the distributions present on Figure \ref{fig:2}, computed using $s_i=\sum_ja_{ij}\times w_{ij}$. So, $p(s_{\beta})$ is the frequency of occurrence of $s_{in}$ (left) and $s_{out}$ (right). The best-fitting offered using equation (\ref{eq:pqw}), the $q$-Weibull distribution (dashed lines), was obtained with the parameters reported on table \ref{tab:par}, fourth and fifth column.}
\end{figure}

To treat with that drop observed on Figure \ref{fig:acum}, we took into account a weight $w_{ij}$ for each service route computing the normalized total number of companies that travel on each route, which are inscribed on the ANTT data-base. It represents giving each route a more realistic importance. Indeed, Figure \ref{fig:Sqw} shows the cumulative distribution corresponding to curves for \emph{in}- and \emph{out}-degree shown on Figure \ref{fig:2}, but where we considered the weight distribution computing $s_i=\sum_ja_{ij}\times w_{ij}$ for each node $i$, and plotting $s\equiv s(k)$ for \emph{in}- and \emph{out}-degree cases. For these weighted cumulative versions, both incoming and outgoing distributions, the best-fitting was obtained following equation (\ref{eq:pqw}). This time there is no drop previously detected on Figure \ref{fig:acum}-$\mathbf{d}$. The insets in Figure \ref{fig:Sqw}$-\mathbf{a}$ show in more details the $q$-Weibull fitting for the head ($\mathbf{b}$) and the tail ($\mathbf{c}$) of the \emph{in}-distribution, whereas insets ($\mathbf{e}$) and ($\mathbf{f}$), on frame $\mathbf{d}$ follow the same organization for the \emph{out}-degree case.

\subsection{More about the power-law description} 
\label{sec:pl}
\begin{figure}[h]
\centering
\includegraphics[scale=0.3]{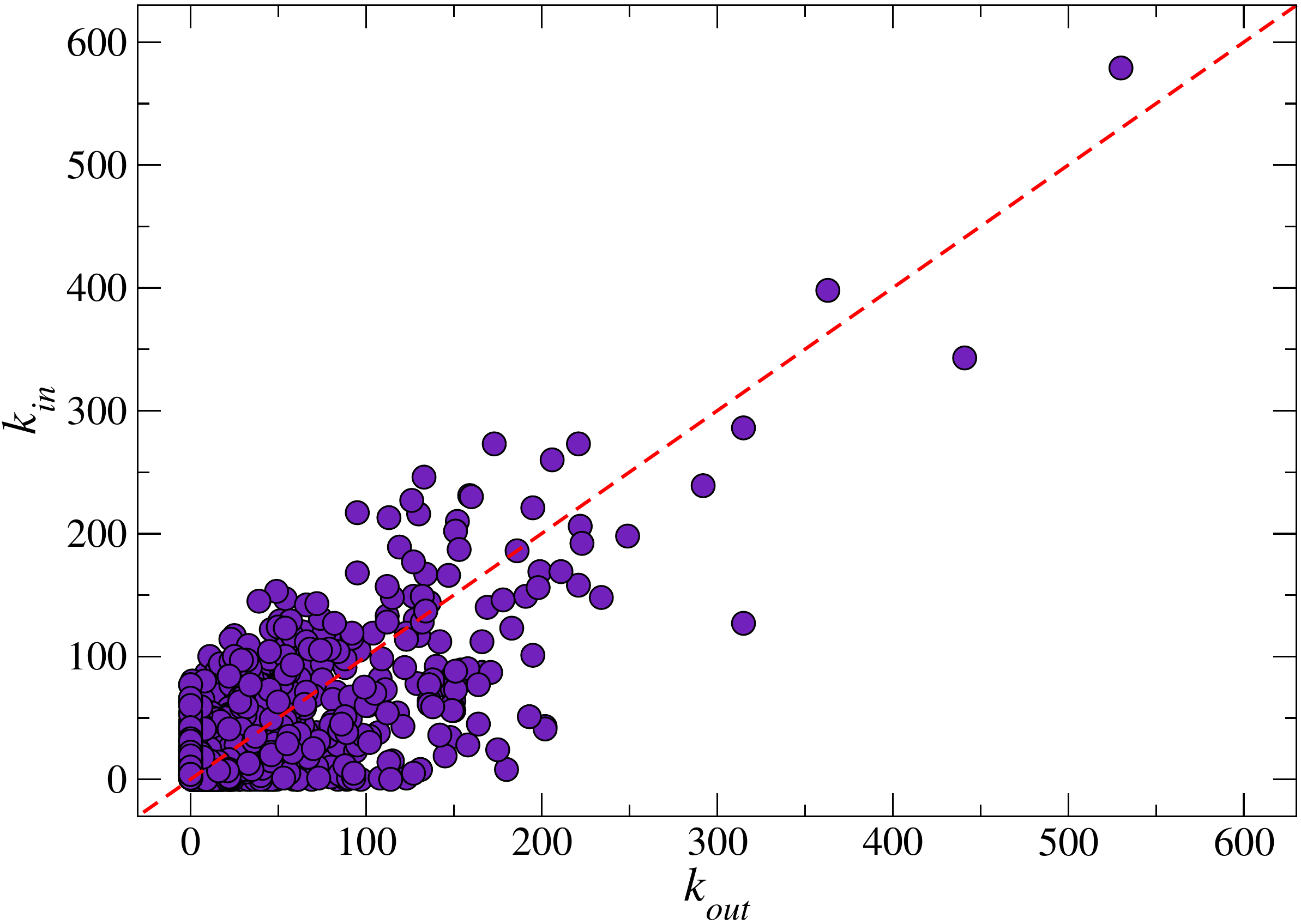}
\caption{\label{fig:pear} \textbf{Linear correlation.} Scatter plot showing the balance between incoming and outgoing connections for each locality. The quantitative Pearson coefficient obtained between the data-set $k_{in}$ $vs.$ $k_{out}$ was $r=+0.76$.}
\end{figure}

We performed a quantitative analyze of the balance between incoming and outgoing connections for each one of the localities present in the network. We carried out the Pearson correlation analyze on the data-set of incoming and outgoing connections, in which the Pearson coefficient $r=+1$ marks a total linear positive correlation. This coefficient of correlation can be obtained by
\begin{equation}
r=\frac{\sum_i^n(k_i^{in}-\langle k^{in}\rangle)(k_i^{out}-\langle k^{out}\rangle)}
{\sqrt{\sum_i^n(k_i^{in}-\langle k^{in}\rangle)^2}\sqrt{\sum_i^n(k_i^{out}-\langle k^{out}\rangle)^2}}.
\label{equation pearson}
\end{equation}
A scatter plot for the investigation of the linear correlation $k_{in}$ \emph{vs.} $k_{out}$ is shown on Figure~\ref{fig:pear} and the result for the Pearson correlation coefficient achieved was $r=+0.76$, meaning a strong correlation. Generally speaking, there is a balanced relation of the number of incoming and outgoing connections for each locality.

Actually, the Pearson coefficient $r=+0.76$ also means that although we are handling with a network with flows, that is, a directed network, the decays of both distributions shown on Figure \ref{fig:2} are roughly the same. So, by supposing  an undirected network, we can address analytically that the often mathematical description for connectivity distributions under road networks should do not correspond to a simple power-law function for the data-set we used. At least, it is not conclusive due the lack of adjustments for the head of the distributions and their great variations on their distribution of moments.

\begin{figure}[h]
\centering
\includegraphics[scale=0.32]{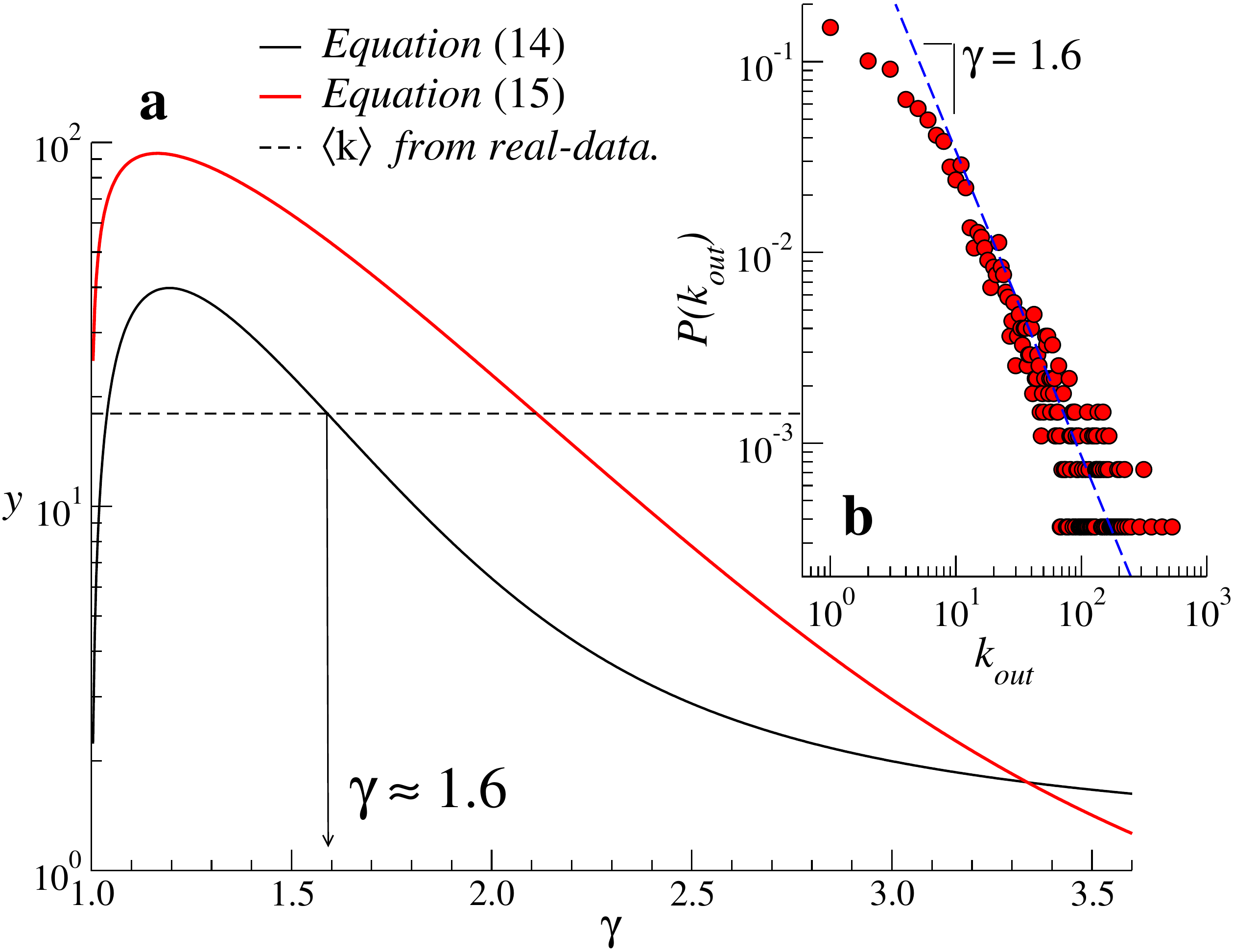}\includegraphics[scale=0.32]{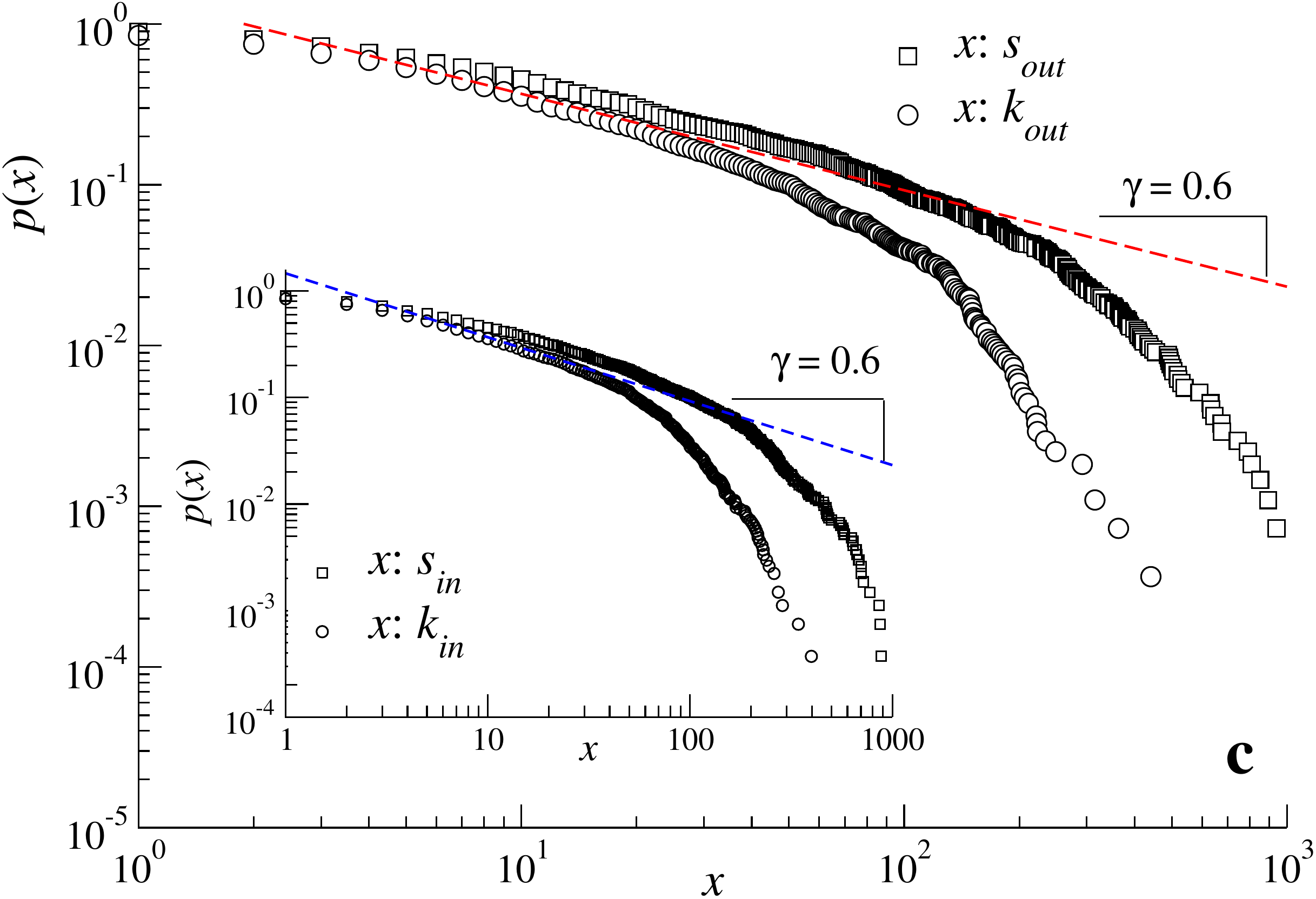}
\caption{\label{fig:theo} \textbf{On the power-law description.} $\mathbf{a}$: vertical axis acts for the mean degree value and variance of $P(k)$ as a function of the degree or scale exponent $\gamma$, supposing a power-law protocol $y(x)\propto x^{-\alpha}$ for the description of the connectivity distribution. From data, the average degree correspond to $\langle k\rangle\approx18$ (dashed line) that leads to a degree exponent $\gamma\approx 1.6$. For this value the associated variance is about $2\langle k\rangle$. $\mathbf{b}$: For the data-set we have used, the power-law description offers $\gamma\approx 1.6$, marked by intersection between the analytical curve for equation (\ref{eq:k2n}) and that from data. The fitting obtained (dashed line) is highlighted in  comparison with data shown in Figure \ref{fig:2}. $\mathbf{c}$: Cumulative distributions $p(k)$ and $p(s)$ from Figures \ref{fig:acum}-$\mathbf{a,d}$ and \ref{fig:Sqw}-$\mathbf{a,d}$, respectively, are compared with the cumulative version (with scale exponent $\gamma=0.6$) of the power-law hypothesis.} 
\end{figure}

In order to achieve some analytical insight concerning on this power-law hypothesis, we consider the following verification protocol: (\emph{i}) We assumed with probability one that we can find at least one value of connection and no longer treat differently a power-law function $y(x)\propto x^{-\alpha}$ and $P(k)=Ck^{-\gamma}$. (\emph{ii}) Moreover, we also understand the proportional constant $C=\left(\int k^{-\gamma}dk\right)^{-1}$ is besides the question, since we require the normalization for $P(k)$, which behaves becomes asymptotically as $P(k)\sim k^{-\gamma}$. Thus, taking the continuous limit of $P(k)$ and remembering that $\langle k\rangle=\sum_kkP(k)$, yields $\langle k^n \rangle\sim m^{\left[(n+1)-\gamma\right]}$, where $m$ is the maximum degree found in the network and $n$ marks the $n^{th}$-moment of the distribution. (\emph{iii}) However, we are of course dealing with real-data and some relevant quantities as the mean value $\langle k\rangle$ and the variance $\sigma^2=\sum_k k^2P(k)-\left(\sum_k kP(k)\right)^2$ of the degree distribution have finite size scales. Thus, we should evaluate the normalization $\int P(k)dk=C\int k^{-\gamma}dk=1$ by means of the term $C=\left(\int k^{-\gamma}dk\right)^{-1}$.
Taking this into account, we can compute a general equation for the moments of the distribution $P(k)$,
\begin{equation}
\langle k^n\rangle=\frac{(\gamma-1)}{(n+1)-\gamma}\left(m^{(n+1)-\gamma}-1\right).
\label{eq:k2n}
\end{equation}
Assuming the minimum degree unitary, the variance on $P(k)$ will be given by
\begin{equation}
\sigma^2=\frac{(\gamma-1)(m^{3-\gamma}-1)}{3-\gamma}-\frac{(\gamma-1)^2(m^{2-\gamma}-1)^2}{(2-\gamma)^2}.
\label{eq:sigma}
\end{equation}
Figure \ref{fig:theo} shows equations (\ref{eq:k2n}) and (\ref{eq:sigma}) as functions of the scale exponent $\gamma$ coming from the power-law description. Since from real-data $\langle k_{out}\rangle\approx\langle k_{in}\rangle\approx18$, we expected that the intersection between this value and the analytical curve for equation (\ref{eq:k2n}) will mark the corresponding scale exponent that well-fits both route distributions. In the Figure \ref{fig:theo}-$\mathbf{b}$ is shown the power-law fitting using the predicted value for the scale exponent $\gamma=1.6$. Although apparently good fit (save initial points) on log-log scale, the cumulative version ($\int^{\infty}_a C x^{\gamma}dx=C'x^{\alpha+1}|_{a}^{\infty}$) of the power-law description fails when compared with those cumulative versions both non-weighted and weighted from Figures \ref{fig:acum}-$\mathbf{a}$,$\mathbf{d}$ and \ref{fig:Sqw}-$\mathbf{a}$,$\mathbf{d}$, respectively, as shown in Figure \ref{fig:theo}-$\mathbf{c}$.

\subsection{Reciprocity coefficient} 
Besides, although the strong Pearson correlation found for the service routes network concerning on the number of incoming and outgoing connections, it remains the question if there are, and at which measure, mutual connections between localities. This analyze was done by performing the ratio between the number of mutual links $L_{m}$ and the total number of links $L$.  However, the reciprocity $\ell=L_{m}/L$ is strongly dependent on the density of links given by
$\bar{a}=\sum a_{ij}/[N(N-1)]=L/[N(N-1)]$
and for this reason have no absolute value for sake of comparison \cite{PhysRevLett.93.268701}. The authors in Ref.~\cite{PhysRevLett.93.268701} proposed an alternative way to compute the reciprocity coefficient:
\begin{equation}
\rho=\frac{\sum_{i\neq j}(a_{ij}-\bar{a})(a_{ji}-\bar{a})}{\sum_{i\neq j}(a_{ij}-\bar{a})^2}=\frac{\ell-\bar{a}}{1-\bar{a}}.
\label{eq:rho}
\end{equation}
Therefore, equation (\ref{eq:rho}) has absolute values such as equation (\ref{equation pearson}). Thus, $\rho=0$ means no reciprocity, $\rho<0$ marks an anti-reciprocity scenario, whereas $\rho>0$ stands for the reciprocal case. For the set of Brazilian service routes we have studied, $\rho=0.2$, meaning a low reciprocity among localities. Actually, this order of magnitude for the reciprocity coefficient was reported in Ref.~\cite{PhysRevLett.93.268701} depicting the mutual links relationship for e-mail network (Address book: $\rho=0.231$; Actual messages: $\rho=0.194$~\cite{PhysRevE.66.035101, PhysRevE.66.035103}). In this kind of network, a message sent by a recipient $A$ to a receiver $B$ are not, in general, answered from $B$ to $A$. 

It is possible to make an analogy of our result to the coefficient of reciprocity $\rho$ with those on Ref. \cite{PhysRevLett.93.268701}: although we expect it have no reality for mobility of passengers, for goods and services (shipping) it may be well described. Such as an e-mail, a cargo can be sent from a recipient $A$ to a receiver $B$ but not necessarily the recipient $B$ will be sent a cargo to $A$. It suggests that there is a disasosiative degree-degree correlation at any level: urban centers (localities with high influence in economy, culture, etc.) with high degree $k_{out}$ send goods to peripherical localities, which in turn does not have much goods to sent to urban centers, or by other way, does not have much localities to express cargoes due to their low $k_{out}$. This disassortative mixture has been shown in frames $\mathbf{c}$ and $\mathbf{d}$ of the Figure \ref{fig:knn}: localities of high $k_{in}$ degree, in the range $k_{in}\gtrsim 60$, are preferentially connected to localities of low $k_{out}$ and vice-versa.

Yet, from the network theory point of view, although the road network we have discussed here is properly understood as a  directed one, it is likely semi-directed:  its reciprocity should not goes to zero as the number of nodes becomes large, but leans to a nonzero constant value~\cite{PhysRevE.66.035101}.

\section*{Discussion}
We have investigated topological features of a road network in Brazil, analyzing a data-set based on the commercial service routes inscribed on the National Land Transport Agency. These routes were represented by links, which in turn interconnect nodes that act as localities. Since there is a flow on this network, it could be studied as a \emph{Origin}-\emph{Destination} matrix, $\mathbf{A}:N\times N$ of entries $\{a_{ij}\}$, and the pattern of incoming and outgoing routes were investigated.

We reported that the connectivity profile of incoming and outgoing connections are well-described by the $q$-Weibull distribution, in a comparison with two different ones, namely the Weibull and $q$-exponential distributions. Interestingly, the outgoing distribution presents a faster drop for large values of $k$, compared to the incoming one. It takes place for a particular range of the number of outgoing routes of the road network. The mechanism responsible for this was investigated by performing a spectral analyze of the eigenvector centrality of each locality. Despite a functional linear relationship between the eigenvector of centrality and the number of outgoing (and incoming) connections is predominant on the range in which this drop takes place, we also observed a drop for this measure of centrality. From a random walk analysis point of view, this indicates the presence of poorly visited places, although high connected. Actually, we have found out that the arrangement of connections inside a neighborhood of each node with degree ranging where the drop occurs are mixed in a disassortative way for both \emph{in}- and \emph{out}- connectivity distributions. However, more pronounced decay were detected for the outgoing case, revealing the mechanism responsible for this. We also took into account the density of companies $w_{ij}$ traveling over each route linking a locality $i$ to another $j$. It correspond to build a weighted road network, which in turn reclaim the real importance of each route.

Moreover, we have performed an analysis using power-law fitting, since this protocol is often used in the road network literature. We derived a mathematical expression for the average number of connections and compare with empirical data to obtained the predicted scale exponent to fit incoming and outgoing distributions. However, we reported that the $q$-Weibull description is the best-fit for the cumulative version (performed aiming to treat statistical fluctuations intrinsic to real-data) of the connectivity distributions with heavy tail we have studied.

Finally, we characterized the linear correlation between the data-set of incoming and outgoing connections and reported the balance of \emph{in}- and \emph{out}- routes for each locality. The corresponding Pearson coefficient obtained was $r=0.76$ meaning a strong linear correlation compared with a fully linear correlated network corresponding to the value $r=1.0$. However, the network does not behave as an undirected one with a fully set of mutual links. In this sense, we have described the reciprocity of the network by obtaining the reciprocal coefficient $\rho=0.2$ (which is far from the limit value $\rho=1.0$ expected for a fully reciprocal case - a fully two-way vias). Although this road network is properly understood as a directed one, it is possibly semi-directed for which its reciprocity does not goes to zero as the number of nodes becomes large, but leans to a nonzero constant value~\cite{PhysRevE.66.035101}. This value is in agreement with that reported for e-mails networks~\cite{PhysRevLett.93.268701} where a message sent by a recipient $A$ to a receiver $B$ is in general one way route. On the other hand, this value is much lower than that reported for the Brazilian airport network ($\rho\approx 0.8$)
\cite{da2009structural}.

Future works could report some investigation concerning on geographic features such the length distributions and temporal evolution of the road network, mainly following the National Logistics Plan schedule (2015-2035) (http://www.epl.gov.br/), as well as for other modals of transport and a multi-modal perspective.

\section*{Acknowledgements}
We specially thank Dr. Antonio de Macedo and Dr. Adam James Sargeant for several suggestions. R.S.F. would
like to thanks the program \textit{Auxílio Pesquisador}/PROPP/UFOP, UESPI/Piripiri/Teresina, Department of
Physics/UFPI.

\end{document}